\pdfoutput=1

\documentclass[11pt]{article}

\usepackage[final]{acl}

\usepackage{times}
\usepackage{latexsym}

\usepackage[T1]{fontenc}

\usepackage[utf8]{inputenc}

\usepackage{microtype}

\usepackage{inconsolata}

\usepackage{graphicx}
\usepackage{xcolor}
\usepackage{pifont} 
\usepackage{amsmath}
\usepackage{amsthm}
\usepackage{booktabs}
\usepackage{comment}
\usepackage{multirow}
\usepackage{subcaption}

\definecolor{myRed}{rgb}{0.808,0.067,0.149}
\definecolor{myGreen}{rgb}{0.067,0.708,0.149}
\title{XMAD-Bench: Cross-Domain Multilingual Audio Deepfake Benchmark}

\author{\bf{Ioan-Paul Ciobanu, Andrei-Iulian Hiji, Nicolae-Catalin Ristea,}\\
{\bf Paul Irofti, Cristian Rusu, Radu Tudor Ionescu$^*$}\\
  Department of Computer Science\\
  University of Bucharest \\
  Bucharest, Romania \\
  $^*$Corresponding author: \texttt{raducu.ionescu@gmail.com}}

\begin{document}
\maketitle

\begin{abstract}
Recent advances in audio generation led to an increasing number of deepfakes, making the general public more vulnerable to financial scams, identity theft, and misinformation. Audio deepfake detectors promise to alleviate this issue, with many recent studies reporting accuracy rates close to $99\%$. However, these methods are typically tested in an in-domain setup, where the deepfake samples from the training and test sets are produced by the same generative models. To this end, we introduce XMAD-Bench, a large-scale cross-domain multilingual audio deepfake benchmark comprising 668.8 hours of real and deepfake speech. In our novel dataset, the speakers, the generative methods, and the real audio sources are distinct across training and test splits. This leads to a challenging cross-domain evaluation setup, where audio deepfake detectors can be tested ``in the wild''. Our in-domain and cross-domain experiments indicate a clear disparity between the in-domain performance of deepfake detectors, which is usually as high as $100\%$, and the cross-domain performance of the same models, which is sometimes similar to random chance. Our benchmark highlights the need for the development of robust audio deepfake detectors, which maintain their generalization capacity across different languages, speakers, generative methods, and data sources. Our benchmark is publicly released at \url{https://github.com/ristea/xmad-bench/}.
\end{abstract}

  \begin{figure}[t]
    \centering
      \includegraphics[width=1.0\linewidth]{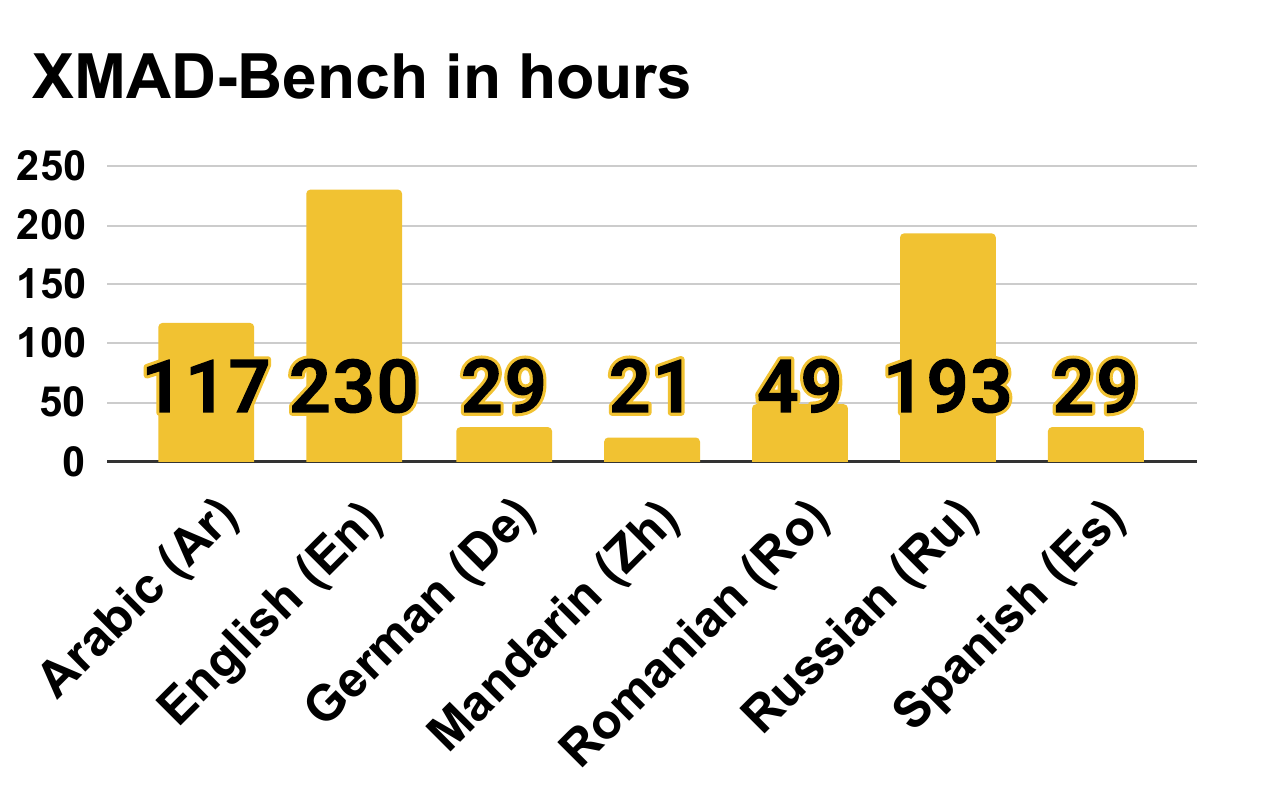}
    \caption{XMAD-Bench comprises 668.8 hours of real and fake speech samples across seven languages: Arabic (Ar), English (En), German (De), Mandarin Chinese (Zh), Romanian (Ro), Russian (Ru), and Spanish (Es). For each language, there are two sources of real samples, enabling us to organize the dataset in a cross-domain format. Best viewed in color.}
    \label{fig_teaser}
  \end{figure}
  
\section{Introduction}

The recent development of powerful audio generation models, capable of synthesizing realistic speech from text \cite{Casanova-ICML-2022,Huang-IJCAI-2023,Ju-ArXiv-2024,Shen-ICLR-2024,Tan-TPAMI-2024} and precisely reproducing voices \cite{Jiang-ArXiv-2023,Lee-ICLR-2023,Wang-ArXiv-2023}, opened the path to new application domains. Unfortunately, these advancements also led to an increase in misuses, especially related to deepfake generation. Indeed, it was recently reported that the number of frauds based on deepfake increased 10-fold from 2022 to 2023\footnote{\href{https://sumsub.com/blog/sumsub-experts-top-kyc-trends-2024/}{Sumsub Expert Roundtable: The Top KYC Trends Coming in 2024}}. This is particularly worrying for the audio domain, since it was found that roughly $70\%$ of the general public is not able to tell if a voice is real or fake\footnote{\href{https://www.mcafee.com/blogs/privacy-identity-protection/artificial-imposters-cybercriminals-turn-to-ai-voice-cloning-for-a-new-breed-of-scam/}{Artificial Imposters--Cybercriminals Turn to AI Voice Cloning for a New Breed of Scam}}. In this context, accurately detecting deepfake audio content is of utter importance. 

To date, considerable research efforts have been dedicated to advance audio deepfake detection, most of the recent approaches being based on deep learning models \cite{Chen-ICASSP-2023,Jung-ICASSP-2022,Liu-ICASSP-2023,tak-ArXiv-2022}. Impressively, such models reached or even surpassed the $99\%$ threshold in terms of audio deepfake detection performance on existing benchmarks \cite{Croitoru-Arxiv-2024}, such as ASVspoof 2019-LA~\cite{wang-CSL-2020} and 
ASVspoof 2021-LA~\cite{yamagishi-ASVspoof-2021}. However, due to the limitations of existing audio deepfake datasets, audio deepfake detectors are generally tested in an in-domain setup, where the deepfake samples from the training and test sets are generated by the same audio generation tools. Hence, the reported accuracy levels of current audio deepfake detectors do not reflect the actual performance of these models when tested ``in the wild'', where the speaker identity or the generative method remains unknown.

To this end, we introduce a novel benchmark for cross-domain audio deepfake detection in multiple languages. XMAD-Bench, which stands for Cross-Domain Multilingual Audio Deepfake Benchmark, contains 668.8 hours of real and fake speech across seven languages: Arabic, English, German, Mandarin Chinese, Romanian, Russian, and Spanish (see Figure \ref{fig_teaser}). XMAD-Bench comprises a variety of languages, including both widely-spoken (English, Arabic) as well as less popular (Romanian) languages. The dataset is balanced in terms of the real versus fake sample distribution, containing 207K real samples and 207K deepfake samples. XMAD-Bench also provides an official three-way split of the data samples into a training set, an in-domain test set and a cross-domain test set, such that speakers are distinct across splits. More importantly, the cross-domain test set contains real audio samples from data sources that are distinct from the training set, and deepfake audio samples generated by a different set of generative methods than the training set. This leads to a challenging cross-domain evaluation setup, which allows audio deepfake detectors to be tested ``in the wild''. Due to the fast pace of AI research, we emphasize that generative methods can become obsolete in 3-4 years, so the ratio of known models in the test set can drastically decrease over time. To take into account the passage of time, we refer to the ``cross-domain'' setting as ``in the wild''. Nevertheless, since we provide both in-domain and cross-domain test sets, interested parties can easily combine samples from the two test sets to obtain an ``in the wild'' setup with a desired ratio. 

We conduct experiments with both convolutional and transformer architectures, namely ResNet-18 \cite{He-CVPR-2015}, ResNet-50 \cite{He-CVPR-2015}, AST \cite{Gong-INTERSPEECH-2021}, SepTr \cite{Ristea-INTERSPEECH-2022}, wav2vec 2.0~\cite{baevski-wav2vec-2020}, and Whisper-Large-v3 \cite{Radford-ICML-2023}. The objective of our experiments is to compare the in-domain and cross-domain performance of neural models with various configurations. Our results show that state-of-the-art models are capable of reaching extremely high accuracy rates (usually close to 100\%) on the in-domain test split, but they fail to maintain their strong performance in the cross-domain setting. The obvious disparity between the in-domain and cross-domain performance of state-of-the-art models indicates that more research efforts need to be dedicated to the development of robust audio deepfake detectors, which maintain their generalization capacity across different speakers, generative methods, and data sources.

In summary, our contribution is twofold:
\begin{itemize}
    \item We introduce XMAD-Bench, a large-scale cross-domain multilingual audio deepfake benchmark comprising 668.8 hours of real and deepfake speech across seven languages.
    \item We carry out comprehensive in-domain and cross-domain experiments to evaluate audio deepfake detectors based on state-of-the-art neural architectures, showing that such models exhibit generally poor generalization capacity.
\end{itemize}

\section{Related Work}

The scientific community uses a relatively small number of existing datasets to assess the effectiveness of audio deepfake detection methods. Such datasets usually contain a single language, which limits their usage in multilingual scenarios, with only a few exceptions, such as MLAAD~\cite{muller2024mlaad} and WaveFake~\cite{frank2021wavefake}, comprising multiple languages for spoofed audio detection. Some of the most popular datasets correspond to the ASVspoof challenges, especially the 2019 and 2021 editions~\cite{wang-CSL-2020, yamagishi-ASVspoof-2021}, which encouraged research in anti-spoofing methods for Automatic Speaker Verification (ASV). 
Both datasets contain only English samples based on the Voice Cloning Toolkit corpus~\cite{vctk2019}.
ADD 2022~\cite{yi2022add} and ADD 2023~\cite{yi2023add} introduce various scenarios, such as low-quality fake audio detection, partially fake audio detection and deepfake algorithm recognition, with a corresponding dataset for each task. They are based on the AISHELL Mandarin speech corpus~\cite{bu2017aishell,Shi-INTERSPEECH-2021,fu2021aishell} and contain fake samples generated with various text-to-speech (TTS) and voice conversion (VC) systems. 
WaveFake~\cite{frank2021wavefake} consists of English and Japanese fake samples generated with different TTS models, starting from real clips from the LJSPEECH~\cite{ito2017lj} and JSUT~\cite{sonobe2017jsut} datasets.
\citet{reimao2019dataset} proposed the Fake or Real (FoR) dataset, with real English clips collected from open-source datasets, such as LJSPEECH~\cite{ito2017lj}, Arctic, VoxForge, and social media platforms such as YouTube. The fake samples are generated using both open-source and commercial TTS systems. 

\begin{table*}[t]
\setlength\tabcolsep{0.22em}
\centering
\small{
\begin{tabular}{llcc cccr}
\toprule
\multirow{4}{*}{\textbf{\rotatebox[origin=c]{90}{Language}}} & \multirow{4}{*}{\textbf{Subset}} & \multirow{4}{*}{\textbf{Data source}} & \multirow{4}{*}{\textbf{Fake generation methods}} &  & \multirow{4}{*}{\textbf{\rotatebox[origin=c]{90}{\#samples}}} &  & \multirow{4}{*}{\textbf{\rotatebox[origin=c]{90}{\#speakers}}} \\
& & & & \multicolumn{1}{c}{\textbf{Total}} & & \multicolumn{1}{c}{\textbf{Average}} & \\
& & & & \textbf{len.~(h)} & & \textbf{len.~(s)} & \\
& & & & & & & \\
\midrule
\multirow{3.5}{*}{\rotatebox[origin=c]{90}{Arabic}} & Training & CommonVoice & fairseq+KNN-VC, fairseq+FreeVC& 68.46& 56,114 & 4.39& 204 \\
 & In-Domain Test & CommonVoice & fairseq+KNN-VC, fairseq+FreeVC& 24.63& 20,488 & 4.32& 52 \\
\cmidrule{2-8}
& Cross-Domain Test & MASC & XTTSv2& 23.80& 12,984 & 6.59& 1,502 \\
\midrule

\multirow{4.5}{*}{\rotatebox[origin=c]{90}{English}} & Training & CommonVoice & VITS+KNN-VC, XTTSv2 & 114.05 & 75,000 & 5.47 & 885 \\
& In-Domain Test & CommonVoice & VITS+KNN-VC, XTTSv2 & 33.99 & 21,368 & 5.73 & 222 \\
\cmidrule{2-8}
& \multirow{2}{*}{Cross-Domain Test} & \multirow{2}{*}{M-AILABS} & VITS+OpenVoice, GlowTTS+FreeVC, & \multirow{2}{*}{82.41} & \multirow{2}{*}{39,690} & \multirow{2}{*}{7.47} & \multirow{2}{*}{3} \\
&  &  & VALL-E-X, YourTTS   &  &  &  &  \\

\midrule
\multirow{4.5}{*}{\rotatebox[origin=c]{90}{German}} & Training & CommonVoice & fairseq+OpenVoice, XTTSv2 & 17.73 & 11,672 & 5.47 & 60 \\
& In-Domain Test & CommonVoice & fairseq+OpenVoice, XTTSv2 & 4.89 & 3,262 & 5.40 & 15 \\
\cmidrule{2-8}
& \multirow{2}{*}{Cross-Domain Test} & \multirow{2}{*}{M-AILABS} & Tacotron2-DDC+FreeVC/KNN-VC, & \multirow{2}{*}{6.78} & \multirow{2}{*}{3,100} & \multirow{2}{*}{7.88} & \multirow{2}{*}{5} \\
&  &  & VITS+FreeVC, VITS+KNN-VC, YourTTS   &  &  &  &  \\
\midrule

\multirow{4.5}{*}{\rotatebox[origin=c]{90}{Mandarin}} & Training & CommonVoice & Tacotron2-DDC-GST+KNN-VC, Bark+FreeVC& 16.56 & 11,766 & 5.06 & 821 \\
& In-Domain Test & CommonVoice & Tacotron2-DDC-GST+KNN-VC, Bark+FreeVC& 2.81 & 2,010 & 5.04 & 45 \\
\cmidrule{2-8}
& \multirow{2}{*}{Cross-Domain Test} & \multirow{2}{*}{AISHELL-3} & MeloTTS+OpenVoice, & \multirow{2}{*}{1.85} & \multirow{2}{*}{2,002} & \multirow{2}{*}{3.33} & \multirow{2}{*}{95} \\
&  &  & VALL-E-X, XTTSv2   &  &  &  &  \\

\midrule
\multirow{3.5}{*}{\rotatebox[origin=c]{90}{Romanian$\!$}} & Training& CommonVoice & VITS+KNN-VC, VITS+FreeVC& 25.20& 25,934 & 3.50& 144 \\
 & In-Domain Test & CommonVoice & VITS+KNN-VC, VITS+FreeVC& 4.72& 4,886 & 3.47& 26 \\
\cmidrule{2-8}
& Cross-Domain Test & VoxPopuli & VITS+OpenVoice& 18.77& 6,672 & 10.13& 38 \\
\midrule

\multirow{3.5}{*}{\rotatebox[origin=c]{90}{Russian}} & Training & CommonVoice & VITS+KNN-VC, XTTSv2 & 86.65 & 56,126 & 5.56 & 158 \\
& In-Domain Test & CommonVoice & VITS+KNN-VC, XTTSv2 & 17.08 & 11,318 & 5.43 & 40 \\
\cmidrule{2-8}
& Cross-Domain Test & M-AILABS & VITS+OpenVoice  & 89.52 & 34,702 & 9.29  & 3 \\
\midrule

\multirow{5.5}{*}{\rotatebox[origin=c]{90}{Spanish}} & Training & CommonVoice & fairseq+OpenVoice, XTTSv2 & 18.75 & 10,436 & 6.47 & 65 \\
& In-Domain Test & CommonVoice & fairseq+OpenVoice, XTTSv2 & 4.05 & 2,258 & 6.47 & 17 \\
\cmidrule{2-8}
& \multirow{3}{*}{Cross-Domain Test} & \multirow{3}{*}{M-AILABS} & MeloTTS+FreeVC, MeloTTS+KNN-VC, & \multirow{3}{*}{6.13} & \multirow{3}{*}{3,070} & \multirow{3}{*}{7.19} & \multirow{3}{*}{3} \\
&  &  & Tacotron2-DDC+FreeVC/KNN-VC,  &  &  &  &  \\
&  &  & VITS+FreeVC, VITS+KNN-VC, YourTTS   &  &  &  &  \\
\midrule

\multirow{3.5}{*}{\rotatebox[origin=c]{90}{Overall}} & Training & - & - & 347.40 & 247,048 & 5.21& 2,337 \\
& In-Domain Test & - & - & 92.17& 65,590 & 5.13 & 417 \\
\cmidrule{2-8}
& Cross-Domain Test & - & -  & 229.26 & 102,220 & 8.23 & 1,649 \\
\bottomrule
\end{tabular}
}
\caption{Statistics for each language and split, as well as the selected generation methods used in each case.}
\label{tab:dataset}
\end{table*}

To evaluate the performance of deepfake detection models in a cross-dataset scenario, \citet{muller2022does} trained several detectors on the ASVspoof2019~\cite{wang-CSL-2020} dataset and reported results on their novel dataset, called MLAAD. They observed a large performance drop, questioning the generalization capability of various deepfake detectors. MLAAD~\cite{muller2024mlaad} includes samples synthesized from the M-AILABS Speech Dataset~\cite{dataset2024m}, being one of the few datasets that contains fake audio samples covering multiple languages. For the majority of languages (not part of M-AILABS), the authors generated fake samples by translating English text samples into additional target languages and then using state-of-the-art TTS models. Despite the large number of languages covered by MLAAD, it contains only fake samples. Therefore, to train a deepfake detector, one also needs a dataset of real samples, containing exactly the same set of languages as MLAAD, but this is not the case for M-AILABS. If the training language sets for real and fake samples are different, the deepfake detector can suffer from significant language biases, i.e.~it can mislabel samples in languages for which the real or fake samples are missing. In contrast, we mitigate this problem by constructing a balanced dataset that includes both real and synthesized samples in all target languages. Moreover, MLAAD itself is not organized to support out-of-domain evaluations, making results reported in different papers hard to compare. In contrast, XMAD-Bench provides a clear organization via an official split that is publicly released, enabling direct comparisons without having to reproduce or retrain models in distinct setups.

A number of concurrent works proposed speech datasets for deepfake detection in multiple languages \cite{Huang-ACL-2025, Sharma-ACL-2025}. SpeechFake \cite{Huang-ACL-2025} offers support in 46 languages, while IndicSynth \cite{Sharma-ACL-2025} covers 12 low-resource Indian languages. Nonetheless, these benchmarks do not specifically address the cross-domain evaluation setup, which enhances the relevance of our contribution.

\citet{Li-EMNLP-2024} identified the generalization issue of deepfake detectors, proposing a monolingual cross-domain dataset for audio deepfake detection, called CD-ADD. The dataset contains 300 hours of speech generated by five zero-shot TTS models, hence the cross-domain nature. The authors also make use of the ASVSpoof2019 dataset, as well as pre-trained speech encoders, such as wav2vec 2.0~\cite{baevski-wav2vec-2020} and Whisper~\cite{Radford-ICML-2023}. The dataset is affected by several perturbations, called ``attacks'', which simulate real-world noise and significantly affect the detection performance. The work does not use any real speech data and does not make an explicit effort to accommodate multiple languages. 

The results obtained by the latest deepfake detection models tend to saturate existing benchmarks, with GNN-based methods, like AASIST~\cite{Jung-ICASSP-2022,tak-ArXiv-2022}, transformer-based methods, like Rawformer~\cite{Liu-ICASSP-2023}, and other models~\cite{Rosello-INTERSPEECH-2023,Truong_2024} reporting EER values of around 1\%. We conjecture that the performance of such models would drop significantly when these models are tested on a dataset that would include audio clips from diverse speakers, recorded in different conditions and with spoofed samples generated by different methods. To the best of our knowledge, XMAD-Bench is the first multilingual cross-domain dataset for audio deepfake detection, containing both real and fake samples. Moreover, XMAD-Bench is the largest dataset of its kind, being more than twice as large as recent large-scale datasets, such as MLAAD~\cite{muller2024mlaad} and CD-ADD~\cite{Li-EMNLP-2024}.

\section{Dataset}

\subsection{Overview}
The XMAD-Bench dataset consists of real audio clips collected from various existing speech datasets, and corresponding fake clips generated based on the real ones. Each real clip has a matching fake version, generated using text-to-speech (TTS) and voice conversion (VC) tools, which preserve the text content and speech characteristics of the original. The dataset includes speech samples in Arabic, English, German, Mandarin Chinese, Romanian, Russian, and Spanish. 
For each language, the clips are sourced from two datasets, one designated for training and in-domain testing, and another for cross-domain testing. The in-domain data is divided into a training split and an in-domain test split. The fake samples for the in-domain data are generated by two distinct generative methods, such that half of the fake audio files are produced by the first method, and the other half by the second method. For the cross-domain test set, fake samples are synthesized by multiple generative methods, distinct from the first two. Moreover, the speakers are distinct across all three splits.

In Table~\ref{tab:dataset}, we list the sources of real samples and the generative methods used for each language. For the in-domain data, the real speech samples are collected from Common Voice~\cite{commonvoice:2020}, a massively-multilingual speech corpus. The real samples for the cross-domain test are gathered from the Massive Arabic Speech Corpus (MASC)~\cite{10022652} (for Arabic), the M-AILABS dataset~\cite{dataset2024m} (for English, German,  Russian and Spanish), the AISHELL-3 corpus~\cite{Shi-INTERSPEECH-2021} (for Mandarin Chinese), and the VoxPopuli dataset~\cite{wang-etal-2021-voxpopuli} (for Romanian). MASC contains audio collected from YouTube in Arabic. AISHELL-3 is a corpus comprising 85 hours of speech from 218 Mandarin speakers. M-AILABS is a multilingual corpus based on audiobooks, while VoxPopuli is a multilingual speech corpus consisting of European Parliament recordings. The datasets are chosen due to their permissive license agreements, which allow us to share data and derivatives for non-commercial research purposes.

\begin{figure}[t]
   \centering
   \includegraphics[width=1\linewidth]{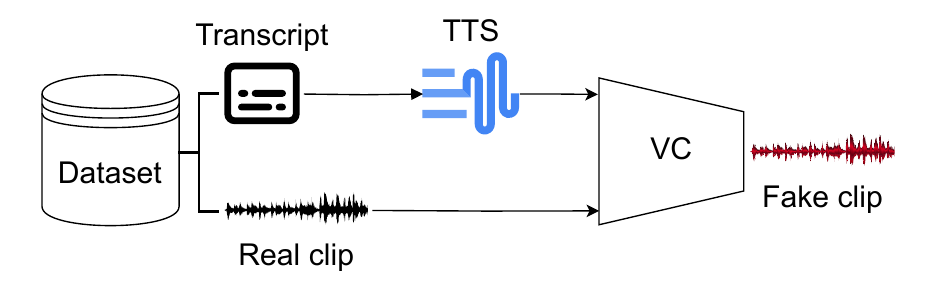}
    \caption{General flow for fake sample generation based on various text-to-speech and voice conversion tools.}
    \label{fig_generation}
\end{figure}

\subsection{Synthesis Procedure}
For fake audio generation, the transcribed text of the real audio sample is first passed to a TTS model. Next, a VC tool is applied to the generated speech, using the voice of the speaker uttering the real sample as reference. The general flow used to generate fake audio samples is  illustrated in Figure \ref{fig_generation}.
For the TTS step, we alternate between various models, namely VITS \cite{kim2021conditionalvariationalautoencoderadversarial},
XTTSv2 \cite{casanova2024xttsmassivelymultilingualzeroshot}, fairseq (based on VITS) \cite{pratap2023mms}, Tacotron2 \cite{Shen-ICASSP-2018}, MeloTTS \cite{Zhao-2024}, YourTTS \cite{Casanova-ICML-2022}, GlowTTS \cite{Kim-NeurIPS-2020}, VALL-E-X \cite{Zhang-ArXiv-2023} and Bark \cite{Bark-2023}.
For the VC step, we alternatively employ KNN-VC \cite{baas2023knnvc},
FreeVC \cite{li2022freevchighqualitytextfreeoneshot} and OpenVoice \cite{qin2024openvoiceversatileinstantvoice}. The TTS and VC models are chosen based on their public availability, state-of-the-art performance and support for the various languages included in XMAD-Bench. Most models, e.g.~VITS, XTTSv2, fairseq, Tacotron2, YourTTS, GlowTTS and Bark, are imported from the Coqui TTS library\footnote{\url{https://github.com/idiap/coqui-ai-TTS}}. Other models, e.g.~KNN-VC, VALL-E-X, MeloTTS, are imported from the official repositories of the corresponding papers. While some models, such as KNN-VC, FreeVC, and OpenVoice, use the clips synthesized by a TTS model as input, XTTSv2, YourTTS and VALL-E-X take both the transcript and the reference voice, and directly generate fake samples, requiring no additional VC tool.
All models are pre-trained on the target languages, requiring no adaptation from our end. Note that fairseq models exhibit lower performance than the other ones, but we decided to keep them due to their multilingual support. 

We employ two alternative synthesis procedures per language for the in-domain data, aiming to increase the variability of the fake samples. The cross-domain test set is generated with as many methods per language as possible, all of them being distinct from those used for the in-domain data.  All the methods extract the speech characteristics from the real samples that are cloned.

For all databases, we employ random sampling to reach the speech lengths reported in Table~\ref{tab:dataset}. For VoxPopuli, in particular, we select clips shorter than 20 seconds, in order to limit variation in clip duration, and discard speakers with less than two minutes of total content duration. The latter threshold is also used to enhance conversion quality when employing KNN-VC. With the exception of KNN-VC, the employed VC tools require a single reference clip. For KNN-VC, multiple clips totaling two minutes are used as reference speech, which improves its VC quality. The real audio samples are typically short (around 5-10 seconds). We generate deepfake samples with equivalent lengths in order to avoid potential spurious correlations between sample lengths and class labels.

The MASC dataset contains many clips from YouTube that vary in length, and may even include music. As music content can be a bias for real clips, we filter clips with a percentage of music content greater than 5\%, using a music detection tool from the TVSM \cite{Hung2022} dataset repository.
As MASC is a massive corpus, only the first 7 minutes from each clip are taken and then split into smaller segments,
each corresponding to a single caption.
After segmentation,
only clips with a duration greater than 4 seconds and a caption shorter than 160 characters are kept, to achieve constancy in clip duration, in concordance with the other datasets.
MASC does not provide speaker identities,
so we simply assume that each speaker appears in only one video. This does not affect the cross-domain nature of XMAD-Bench, since the speakers in CommonVoice and MASC are different.

Finally, all real and fake clips are trimmed for silence at both ends, and resampled to 16 \texttt{kHz}. This ensures that real and fake samples cannot be distinguished based on the sampling rate or the presence of silence periods.

\subsection{Statistics}
XMAD-Bench is composed of 668.8 hours of real and fake content coming from 4,403 different speakers. As shown in Table~\ref{tab:dataset},
there is a noticeable variation in average clip duration and speaker diversity, across domains and source datasets. Since the M-AILABS dataset is composed of audiobooks, its speaker diversity is quite low. The average length of audio clips in VoxPopuli is generally longer, even after we discarded clips longer than 20 seconds. However, these differences do not introduce any spurious correlations in the dataset, since the training split is consistent across languages.

\begin{table}[t!]
\setlength\tabcolsep{0.4em}
\centering
\begin{tabular}{lccccc}
\toprule
 \textbf{Class} & \textbf{SAR} & \textbf{SNR} & \textbf{SIG} & \textbf{BAK} & \textbf{OVRL} \\
\midrule
\textbf{Real} & 0.8563 & 34.54 & 3.17 & 3.96 & 3.55 \\
\textbf{Fake} &	0.8894 & 34.94 & 3.12 & 4.16 & 3.65 \\
\bottomrule
\end{tabular}
\caption{SAR, SNR, SIG, BAK and OVRL scores for real and fake samples from our dataset. Perceptual quality metrics are computed with the model proposed by \citet{Reddy-ICASSP-2022}.}
\label{tab:sar-snr}
\end{table}

To assess the signal quality differences between real and fake samples, we report the speech activity ratio (SAR) and the signal-to-noise ratio (SNR). To assess the difference between real and fake samples in terms of perceived quality, we employ a perceptual evaluation model \cite{Reddy-ICASSP-2022}. This model produces three perceptual scores: quality of speech (SIG), background noise (BAK), and overall quality (OVRL). We report all quality metrics for both real and fake samples in Table~\ref{tab:sar-snr}. We highlight that the SAR and SNR values are fairly similar for both real and fake samples, indicating that the fake samples are of high quality. The SIG, BAK and OVRL values further confirm that the perceptual quality of deepfake samples matches that of real samples, which is consistent with our previous observation based on SAR and SNR. 

\section{Experiments}

\subsection{Audio Deepfake Detectors}
For deepfake detection, we consider four pre-trained models, namely ResNet-18 and ResNet-50 \cite{He-CVPR-2015} from Torchvision\footnote{\url{https://github.com/pytorch/vision}}, Audio Spectrogram Transformer (AST) \cite{Gong-INTERSPEECH-2021} from Hugging Face\footnote{\url{https://huggingface.co/MIT/ast-finetuned-audioset-10-10-0.4593}}, and wav2vec 2.0 \cite{baevski-wav2vec-2020,tak-ArXiv-2022} from Hugging Face\footnote{\url{https://huggingface.co/docs/transformers/model_doc/wav2vec2}}. We fine-tune these models on XMAD-Bench. We also train a fifth model from scratch, namely SepTr \cite{Ristea-INTERSPEECH-2022}. The SepTr model\footnote{\url{https://github.com/ristea/septr}} is composed of 3 separable transformer blocks, each with 5 attention heads. The dimension of each head and the dimension of the MLP layer are set to 256. Each audio clip is converted into a spectrogram to be processed by each of the five models.

\citet{Luo-HCI-2024} found that using a frozen Whisper encoder leads to robust cross-domain performance in deepfake detection. To this end, we introduce another baseline that uses a frozen Whisper-Large-v3 encoder \cite{Radford-ICML-2023} to extract audio features. For each audio sample, we apply Global Average Pooling in the time domain to aggregate the extracted features into a single embedding. The resulting embeddings are used to fine-tune a shallow Multi-Layer Perceptron (MLP).

\subsection{Experimental Setup}
We conduct experiments by training detectors on the training set of each language, and evaluating them on the in-domain test set, after every epoch. The checkpoints achieving the highest performance on the in-domain split are further tested on the cross-domain test set. In addition, we also carry out cross-lingual experiments, training the models on Arabic, German, Romanian, Russian and Spanish, and testing them on English and Mandarin. In the cross-lingual setup, we randomly select at most 3,000 samples per language. In all experiments, each clip is augmented during training with a probability of $0.5$. Augmentations include time shifting by rolling the signal, speed augmentation, volume augmentation by applying random gain, clipping, reverberation, spectral shifting (using high-shelf, low-shelf and peak filters), and pitch shifting.

\begin{table*}[th!]
\setlength\tabcolsep{0.32em}
\centering
\small{
\begin{tabular}{llcccccc}
\toprule
\multirow{4}{*}{\textbf{\rotatebox[origin=c]{90}{Language}}} &
\multirow{4}{*}{\textbf{Method}} & \multicolumn{3}{c}{\multirow{2}{*}{\textbf{In-Domain}}} & \multicolumn{3}{c}{\multirow{2}{*}{\textbf{Cross-Domain}}} \\
& & & & & & & \\
& & \multirow{2}{*}{\textbf{ACC}$\,\uparrow$} & \multirow{2}{*}{\textbf{AUC}$\,\uparrow$} & \multirow{2}{*}{\textbf{EER}$\,\downarrow$} & \multirow{2}{*}{\textbf{ACC}$\,\uparrow$} & \multirow{2}{*}{\textbf{AUC}$\,\uparrow$} & \multirow{2}{*}{\textbf{EER}$\,\downarrow$} \\
& & & & & & & \\

\midrule
\multirow{6}{*}{\rotatebox[origin=c]{90}{Arabic}}
& ResNet-18& 100.0 ($\pm$~0.00)& 100.0 ($\pm$~0.00)& 0.00 ($\pm$~0.00)& 37.80 ($\pm$10.11)& 41.26 ($\pm$12.02)& 57.26 ($\pm$10.29)\\
& ResNet-50& 100.0 ($\pm$~0.00)& 100.0 ($\pm$~0.00)& 0.00 ($\pm$~0.00)& 25.09 ($\pm\,$~4.03)& 29.67 ($\pm\,$~2.37)& 68.50 ($\pm\,$~2.34)\\
& SepTr& 96.84 ($\pm$~1.32)& 98.66 ($\pm$~1.09)& 5.04 ($\pm$~2.61)& 36.15 ($\pm\,$~8.44)& 24.88 ($\pm\,$~6.32)& 60.56 ($\pm\,$~4.96)\\
& AST& 99.97 ($\pm$~0.01)& 99.99 ($\pm$~0.00)& 0.01 ($\pm$~0.01)& 73.39 ($\pm\,$~1.54)& 81.04 ($\pm\,$~0.02)& 26.68 ($\pm\,$~0.23)\\
& wav2vec 2.0 & 100.0 ($\pm$~0.00) & 100.0 ($\pm$~0.00) & 0.00 ($\pm$~0.00) & 75.03 ($\pm\,$~2.11) & 83.27 ($\pm\,$~0.18) & 25.81 ($\pm\,$~0.54)\\
& Whisper+MLP & 99.99 ($\pm$~0.00) &	100.0 ($\pm$~0.00) &	0.00 ($\pm$~0.00) &	53.17 ($\pm$~0.82) &	66.46 ($\pm$~1.40) &	37.74 ($\pm$~1.14) \\
\midrule

\multirow{6}{*}{\rotatebox[origin=c]{90}{English}}
& ResNet-18 & 100.0 ($\pm$~0.00) & 99.99 ($\pm$~0.01) & $\,\,$0.00 ($\pm$~0.00) & 47.76 ($\pm$~3.01) & 47.34 ($\pm\,$~8.84) & 56.63 ($\pm$~7.99) \\
& ResNet-50 & 100.0 ($\pm$~0.00) & 100.0 ($\pm$~0.00) & $\,\,$0.00 ($\pm$~0.00) & 54.02 ($\pm$~9.64) & 59.56 ($\pm$11.94) & 45.53 ($\pm$~9.31) \\
& SepTr & 93.21 ($\pm$~0.79) & 94.67 ($\pm$~2.73) & 12.29 ($\pm$~3.87) & 42.30 ($\pm$~8.66) & 39.20 ($\pm$16.01) & 59.52 ($\pm$12.04) \\ 
& AST & 99.20 ($\pm$~0.32) & 99.96 ($\pm$~0.02)& $\,\,$0.66 ($\pm$~0.17) & 69.19 ($\pm$~1.48) & 79.70 ($\pm\,$~1.78) & 28.52 ($\pm\,$~1.45) \\
& wav2vec 2.0 & 100.0 ($\pm$~0.00) & 100.0 ($\pm$~0.00) & $\,\,$0.00 ($\pm$~0.00) & 69.67 ($\pm$~3.22) & 80.03 ($\pm\,$~1.72) & 28.77 ($\pm\,$~1.55)\\
& Whisper+MLP & 99.91 ($\pm$~0.02) &	99.99 ($\pm$~0.00) &	$\,\,$0.10 ($\pm$~0.02) &	90.94 ($\pm$~1.11) &	97.31 ($\pm$~0.82) &	$\,\,$8.38 ($\pm\,$~1.57) \\
\midrule

\multirow{6}{*}{\rotatebox[origin=c]{90}{German}}
& ResNet-18& 100.0 ($\pm$~0.00)& 100.0 ($\pm$~0.00)& 0.00 ($\pm$~0.00)& 90.31 ($\pm$~6.08)& 99.49 ($\pm$~0.34)& $\,\,$3.65 ($\pm$~1.54)\\
& ResNet-50& 100.0 ($\pm$~0.00)& 100.0 ($\pm$~0.00)& 0.00 ($\pm$~0.00)& 96.96 ($\pm$~1.09)& 99.79 ($\pm$~0.02)& $\,\,$2.16 ($\pm$~0.15)\\
& SepTr& 99.95 ($\pm$~0.43)& 99.25 ($\pm$~0.22)& 3.82 ($\pm$~0.71)& 79.62 ($\pm$~1.28)& 86.54 ($\pm$~2.19)& 19.95 ($\pm$~1.32)\\
& AST& 99.65 ($\pm$~0.09)& 99.99 ($\pm$~0.00)& 0.32 ($\pm$~0.07)& 91.31 ($\pm$~2.40)& 97.50 ($\pm$~1.03)& $\,\,$8.68 ($\pm$~2.27)\\
& wav2vec 2.0 & 100.0 ($\pm$~0.00) & 100.0 ($\pm$~0.00) & 0.00 ($\pm$~0.00) & 94.94 ($\pm$~2.03) & 98.29 ($\pm$~1.65) & $\,\,$3.72 ($\pm$~0.67) \\
& Whisper+MLP & 99.71 ($\pm$~0.01) &	99.99 ($\pm$~0.00) &	0.33 ($\pm$~0.03) &	72.84 ($\pm$~1.09) &	91.01 ($\pm$~0.19) &	16.44 ($\pm$~0.48) \\
\midrule

\multirow{6}{*}{\rotatebox[origin=c]{90}{Mandarin}}
& ResNet-18& 100.0 ($\pm$~0.00)& 100.0 ($\pm$~0.00)& 0.00 ($\pm$~0.00)& 68.79 ($\pm$~3.09)& 87.08 ($\pm$~3.16)& 21.30 ($\pm$~3.63)\\
& ResNet-50& 100.0 ($\pm$~0.00)& 100.0 ($\pm$~0.00)& 0.00 ($\pm$~0.00)& 67.31 ($\pm$~2.81)& 87.55 ($\pm$~1.39)& 20.50 ($\pm$~1.63)\\
& SepTr& 96.14 ($\pm$~1.44)& 99.07 ($\pm$~0.83)& 4.30 ($\pm$~2.35)& 65.10 ($\pm$~4.83)& 78.82 ($\pm$~2.59)& 28.50 ($\pm$~2.60)\\
& AST& 99.28 ($\pm$~0.20)& 99.95 ($\pm$~0.02)& 0.82 ($\pm$~0.28)& 77.52 ($\pm$~1.10)& 84.77 ($\pm$~0.66)& 23.20 ($\pm$~1.02)\\
& wav2vec 2.0 & 99.97 ($\pm$~0.01) & 99.99 ($\pm$~0.00) & 0.01 ($\pm$~0.01) & 71.04 ($\pm$~2.71) & 83.78 ($\pm$~2.01) & 25.43 ($\pm$~1.98) \\
& Whisper+MLP & 99.99 ($\pm$~0.01) &	100.0 ($\pm$~0.00) &	0.01 ($\pm$~0.01) &	76.67 ($\pm$~0.56) &	88.89 ($\pm$~0.32) &	19.01 ($\pm$~0.38) \\
\midrule

\multirow{6}{*}{\rotatebox[origin=c]{90}{Romanian}}
& ResNet-18& 100.0 ($\pm$~0.00)& 100.0 ($\pm$~0.00)& 0.00 ($\pm$~0.00)& 61.59 ($\pm\,$~4.96)& 96.32 ($\pm\,$~1.37)& $\,\,$8.53 ($\pm\,$~2.32)\\
& ResNet-50& 100.0 ($\pm$~0.00)& 100.0 ($\pm$~0.00)& 0.00 ($\pm$~0.00)& 76.39 ($\pm$10.48)& 93.16 ($\pm\,$~2.04)& 13.83 ($\pm\,$~2.30)\\
& SepTr& 98.51 ($\pm$~1.61)& 98.55 ($\pm$~2.00)& 3.96 ($\pm$~4.62)& 48.31 ($\pm\,$~1.56)& 45.98 ($\pm$16.68)& 52.55 ($\pm$12.43)\\
& AST& 99.32 ($\pm$~0.39)& 99.98 ($\pm$~0.01)& 0.43 ($\pm$~0.21)& 57.18 ($\pm\,$~3.76)& 92.60 ($\pm\,$~2.07)& 13.25 ($\pm\,$~3.22)\\
& wav2vec 2.0 & 100.0 ($\pm$~0.00) & 100.0 ($\pm$~0.00) & 0.00 ($\pm$~0.00) & 78.12 ($\pm\,$~3.21) & 93.21 ($\pm\,$~1.99) & 12.09 ($\pm\,$~1.39) \\
& Whisper+MLP & 99.92 ($\pm$~0.03) &	99.99 ($\pm$~0.00) &	0.06 ($\pm$~0.02) &	75.72 ($\pm$~3.24) &	99.54 ($\pm$~0.32) &	$\,\,$2.17 ($\pm\,$~1.56) \\
\midrule

\multirow{6}{*}{\rotatebox[origin=c]{90}{Russian}}
& ResNet-18 & 100.0 ($\pm$~0.00) & 100.0 ($\pm\,$~0.00) & $\,\,$0.00 ($\pm\,$~0.00) & 47.67 ($\pm\,$~3.72) & 36.70 ($\pm\,$~1.99) & 60.38 ($\pm\,$~1.25) \\
& ResNet-50 & 100.0 ($\pm$~0.00) & 100.0 ($\pm\,$~0.00) & $\,\,$0.00 ($\pm\,$~0.00) & 33.85 ($\pm$11.35) & 25.99 ($\pm$16.72) & 69.11 ($\pm$12.77) \\
& SepTr & 94.96 ($\pm$~1.08) & 82.29 ($\pm$22.36) & 20.83 ($\pm$20.38) & 29.22 ($\pm\,$~3.36) & 52.35 ($\pm$20.57) & 49.43 ($\pm$15.45) \\
& AST & 99.69 ($\pm$~0.16) & 99.99 ($\pm\,$~0.05) & $\,\,$0.30 ($\pm\,$~0.15) & 66.24 ($\pm\,$~2.67) & 74.07 ($\pm\,$~3.92) & 32.26 ($\pm\,$~3.65)  \\
& wav2vec 2.0 & 100.0 ($\pm\,$~0.00) & 100.0 ($\pm\,$~0.00) & $\,\,$0.00 ($\pm\,$~0.00) & 61.79 ($\pm\,$~3.92) & 68.54 ($\pm\,$~2.03) & 35.66 ($\pm\,$~3.78) \\
& Whisper+MLP & 99.95 ($\pm$~0.01) &	99.99 ($\pm$~0.00) &	0.05 ($\pm$~0.02) &	93.92 ($\pm\,$~1.53) &	99.74 ($\pm\,$~0.07) &	$\,\,$2.22 ($\pm\,$~0.35) \\
\midrule

\multirow{6}{*}{\rotatebox[origin=c]{90}{Spanish}}
& ResNet-18& 100.0 ($\pm$~0.00)& 100.0 ($\pm$~0.00)& 0.00 ($\pm$~0.00)& 90.21 ($\pm$~1.68)& 98.78 ($\pm\,$~0.56)& $\,\,$4.90 ($\pm\,$~1.19)\\
& ResNet-50& 100.0 ($\pm$~0.00)& 100.0 ($\pm$~0.00)& 0.00 ($\pm$~0.00)& 89.72 ($\pm$~2.74)& 99.27 ($\pm\,$~0.22)& $\,\,$4.07 ($\pm\,$~0.58)\\
& SepTr& 96.47 ($\pm$~0.59)& 98.98 ($\pm$~0.61)& 4.78 ($\pm$~1.99)& 75.22 ($\pm$~8.58)& 79.19 ($\pm$14.33)& 24.52 ($\pm$12.22)\\
& AST& 98.80 ($\pm$~0.28)& 99.95 ($\pm$~0.03)& 0.97 ($\pm$~0.33)& 93.77 ($\pm$~1.60)& 98.68 ($\pm\,$~0.59)& $\,\,$5.73 ($\pm\,$~1.70)\\
& wav2vec 2.0 & 99.97 ($\pm$~0.01) & 99.99 ($\pm$~0.00) & 0.01 ($\pm$~0.01) & 94.62 ($\pm$~1.44) & 98.87 ($\pm\,$~0.38) & $\,\,$4.76~($\pm\,$~0.78) \\
& Whisper+MLP & 99.88 ($\pm$~0.03) &	99.99 ($\pm$~0.00) &	0.12 ($\pm$~0.05) &	71.06 ($\pm$~0.38) &	96.12 ($\pm\,$~0.42) &	$\,\,$8.37 ($\pm\,$~0.53) \\
\midrule

\multirow{6}{*}{\rotatebox[origin=c]{90}{Cross-lingual}}
& ResNet-18& 99.98 ($\pm$~0.01)& 99.99 ($\pm\,$~0.00)& $\,\,$0.01 ($\pm\,$~0.01)& 87.87 ($\pm$~0.99)& 94.41 ($\pm\,$~1.37)& 12.52 ($\pm\,$~1.32)\\

& ResNet-50& 99.96 ($\pm$~0.01)& 99.99 ($\pm\,$~0.00)& $\,\,$0.02 ($\pm\,$~0.01)& 86.10 ($\pm$~2.64)& 93.51 ($\pm\,$~1.30)& 13.67 ($\pm\,$~1.78)\\

& SepTr & 91.70 ($\pm$~4.26)& 82.76 ($\pm$15.66)& 22.05 ($\pm$15.92)& 72.67 ($\pm$~4.22)& 55.87 ($\pm$13.06)& 45.69 ($\pm$10.24)\\

& AST & 98.53 ($\pm$~0.46)& 99.87 ($\pm\,$~0.06)& $\,\,$1.35 ($\pm\,$~0.42)& 81.61 ($\pm$~0.70)& 89.77 ($\pm\,$~1.36)& 16.91 ($\pm\,$~0.91)\\

& wav2vec 2.0 & 99.97 ($\pm$~0.01) & 99.99 ($\pm\,$~0.00) & $\,\,$0.01 ($\pm\,$~0.01) & 89.07 ($\pm$~0.56) & 93.58 ($\pm\,$~1.16) & 11.82 ($\pm\,$~1.07) \\

& Whisper+MLP & 99.75 ($\pm$~0.02) &	99.99 ($\pm$~0.00) &	0.24 ($\pm$~0.01) &	72.80 ($\pm$~0.67) &	90.22 ($\pm\,$~0.79) &	17.77 ($\pm\,$~1.10) \\
\bottomrule
\end{tabular}
}
\caption{Results obtained for both in-domain and cross-domain scenarios for each language, as well as for a cross-lingual setup. We report the average and the corresponding standard deviation for the accuracy (ACC), the area under the curve (AUC), and the equal error rate (EER), over three runs. The symbols $\uparrow$ and $\downarrow$ indicate that upper or lower values are better, respectively.}
\label{tab:results}
\end{table*}

\begin{figure}[!t]
\begin{subfigure}[t]{.48\linewidth}
  \centering
  \includegraphics[width=.99\linewidth]{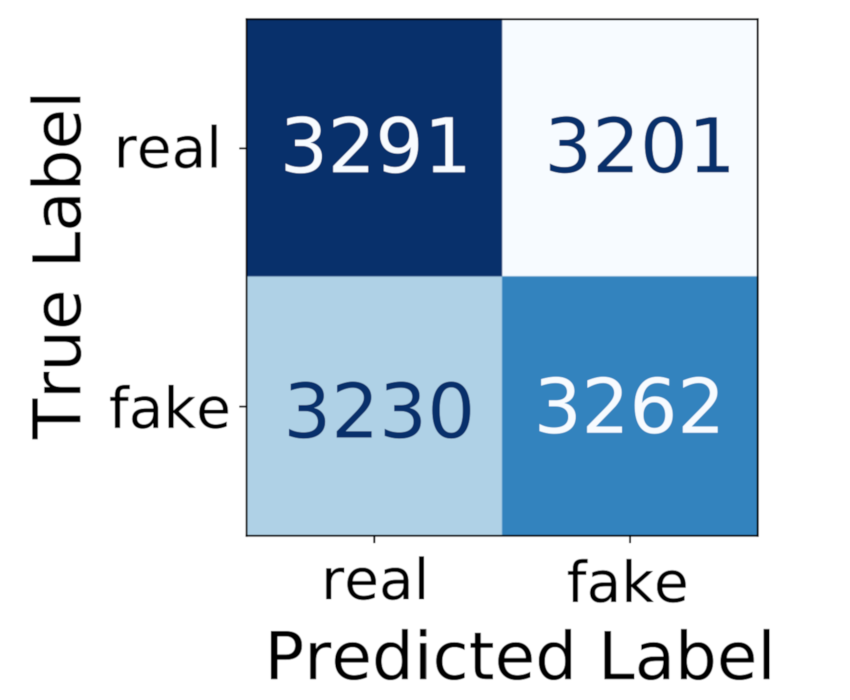}  
  \vspace{-0.5cm}
  \caption{ResNet-18 on Arabic.}
  \label{fig:cm-rn_ar}
    \vspace{0.2cm}
\end{subfigure}
\hfill
\begin{subfigure}[t]{.48\linewidth}
  \centering
\includegraphics[width=.99\linewidth]{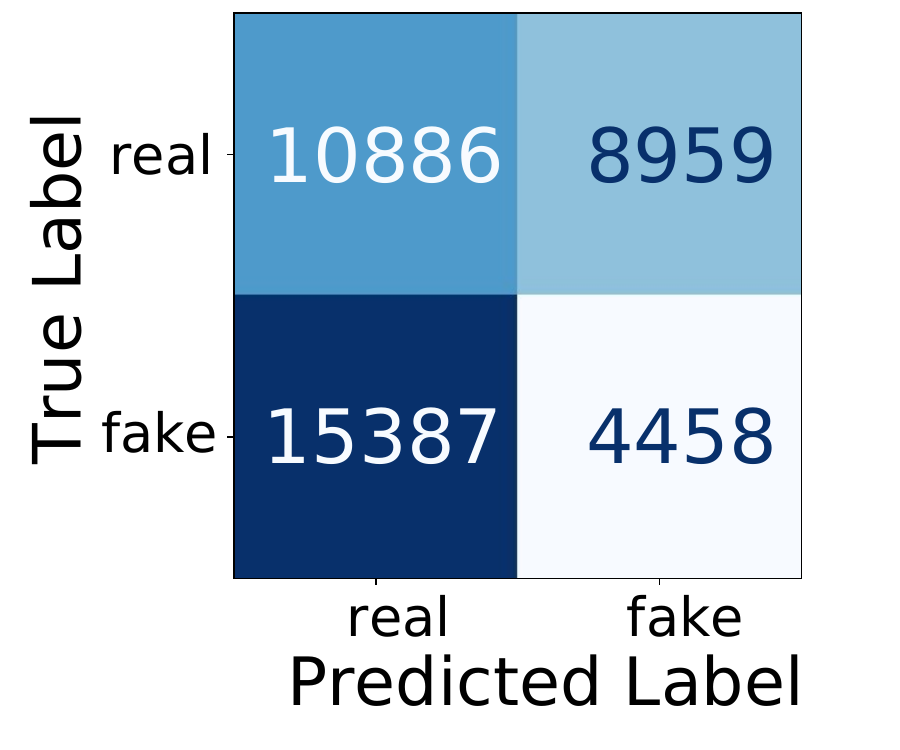}  
\vspace{-0.5cm}
  \caption{ResNet-18 on English.}
  \label{fig:cm-rn_en}
  \vspace{0.2cm}
\end{subfigure}
\begin{subfigure}[t]{.48\linewidth}
  \centering
  \includegraphics[width=.99\linewidth]{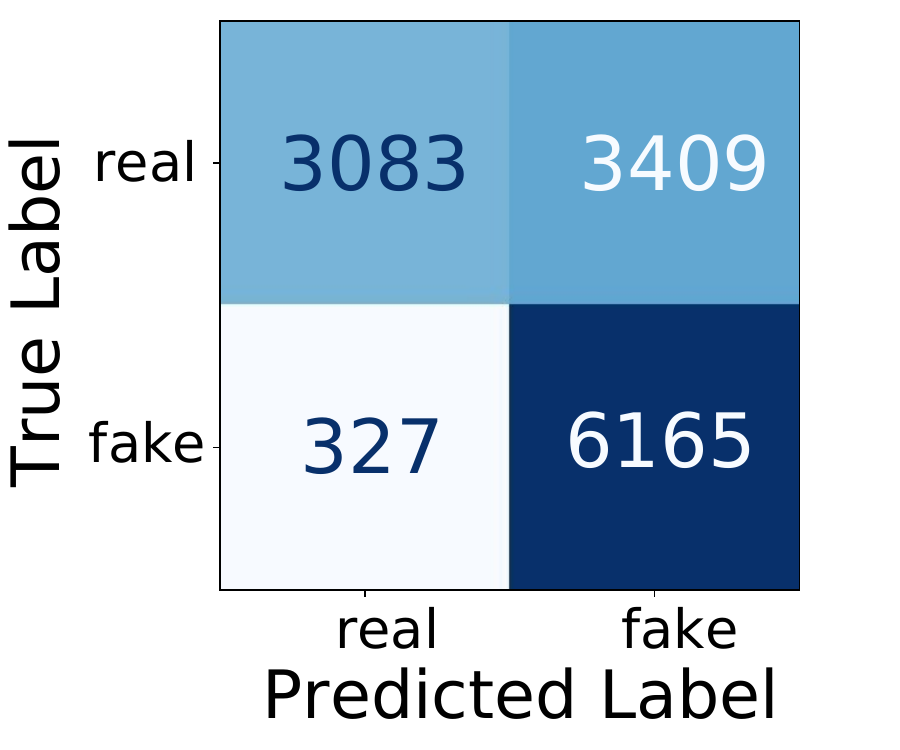}  
  \vspace{-0.5cm}
  \caption{AST on Arabic.}
  \label{fig:cm-ast_ar}
\end{subfigure}
\hfill
\begin{subfigure}[t]{.48\linewidth}
  \centering
\includegraphics[width=.99\linewidth]{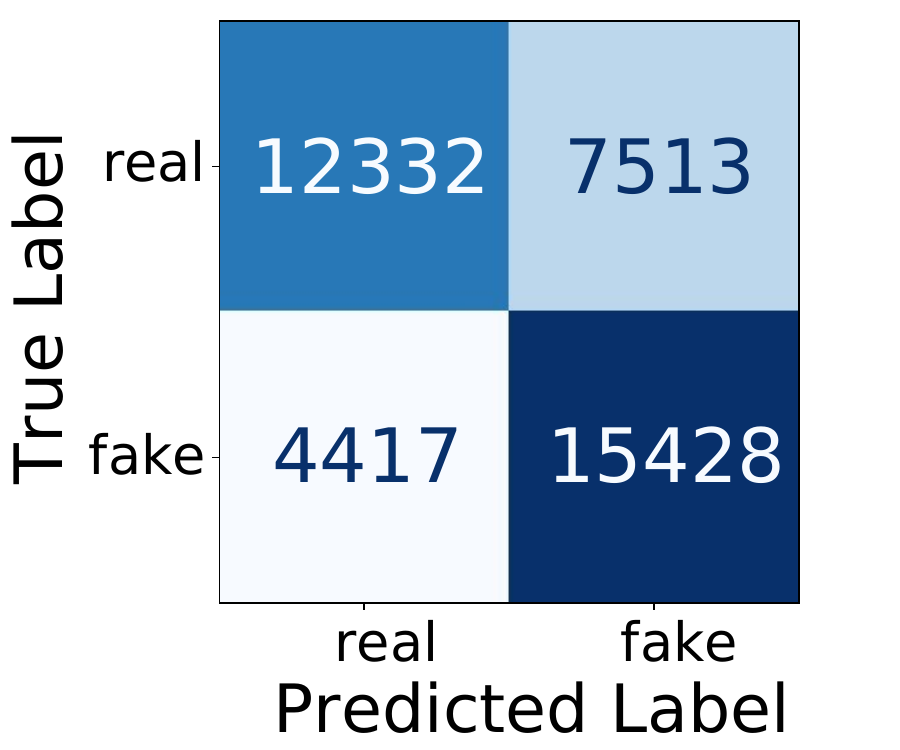} 
  \vspace{-0.5cm}
  \caption{AST on English.}
  \label{fig:cm-ast_en}
\end{subfigure}
\caption{Cross-domain confusion matrices of ResNet-18 (first row) and AST (second row) on Arabic (first column) and English (second column). Best viewed in color.}
\label{fig_confusion_all}
\end{figure}

\subsection{Hyperparameters}
We optimize all models via the cross-entropy loss. Each model is trained for 20 epochs, using a learning rate of $5\cdot 10^{-4}$ and no weight decay. The mini-batch size for each model depends on the size of the respective model. We thus set the mini-batch size to 200 for ResNet-18, 120 for ResNet-50, 16 for wav2vec 2.0, and 10 for both AST and SepTr. The input length of all models is fixed to 5 seconds. We randomly select a segment of 5 seconds from longer clips, while zero-padding the shorter ones. The spectrograms are generated using a 320-point Short-Time Fourier Transform, with 160 hops in the time-domain, on which we apply a Hann sliding window. The resulting size of a spectrogram is $499\times161$. For efficiency reasons, we downsample the spectrogram by a factor of 4 for SepTr. Unlike the other models, AST is based on Mel spectrograms of $1024\times128$ components, this being the default configuration for AST. The wav2vec 2.0 model directly consumes raw audio waveforms sampled at 16 kHz, requiring no handcrafted spectral preprocessing.

\subsection{Main Results}
As shown in Table \ref{tab:results}, several models reach an impressive performance of 100\% on the in-domain split, confirming that deep neural models can easily capture the characteristics of generative models, regardless of the target language. However, the cross-domain results support our conjecture, demonstrating that most metrics drastically decrease when detectors are tested in the cross-domain scenario, where fake samples are generated by models unknown to the detectors. Notably, wav2vec 2.0 and Whishper+MLP outperform all other models in the cross-domain setting on most target languages, both showing a strong ability to generalize to unseen generative models. Furthermore, wav2vec 2.0 exhibits robust multilingual performance, ranking among the top performers in the multilingual setting and demonstrating its effectiveness in capturing language-independent representations of synthetic speech. The cross-lingual in-domain experiments indicate that the language shift does not significantly affect performance. The performance still degrades in the cross-lingual cross-domain scenario.

In Figure \ref{fig_confusion_all}, we present the confusion matrices for the cross-domain evaluation of ResNet-18 and AST on Arabic and English, respectively. On Arabic,
AST tends to label many real samples as fake, while ResNet-18 exhibits both types of mistakes. On English, the two models have opposite biases. Overall, the confusion matrices indicate that the biases are specific to the models, not to XMAD-Bench. This observation further confirms that XMAD-Bench is a challenging and robust benchmark.



\subsection{Results for Background Noise Injection}

We further evaluate how one of the best performing models handles the deepfake detection task under background noise injection. More specifically, we employ the Whisper-based model on Romanian test samples augmented with Gaussian noise, using two values for the standard deviation of the added noise, namely $\sigma=0.01$ and $\sigma=0.1$. Both noise levels can be heard by humans, but the second one makes it difficult to distinguish some words. 

In Table \ref{tab:noise}, we show the cross-domain results of the Whisper+MLP model trained on clean samples (without noise injection). The results indicate that the introduction of audible yet moderate noise ($\sigma=0.01$) does not affect performance. Increasing the noise magnitude ($\sigma=0.1$) degrades performance, but at this noise level, some utterances are indistinguishable by humans. Overall, we conclude that Whisper+MLP obtains reasonable performance when samples are affected by noise, likely because the pre-trained Whisper-Large-v3 backbone is robust to background noise injection.

\begin{table}[t]
\centering
\begin{tabular}{lccc}
\toprule
\textbf{Noise level} & \textbf{ACC}$\,\uparrow$ & \textbf{AUC}$\,\uparrow$ & \textbf{EER}$\,\downarrow$ \\
\midrule
$\sigma = 0$ (no noise) & 75.72 & 99.54 & 2.17 \\
$\sigma = 0.01$ & 80.77 & 98.78 & 4.64 \\
$\sigma = 0.1$ & 68.70 & 98.29 & 5.93 \\
\bottomrule
\end{tabular}
\caption{Cross-domain results of the Whisper-based method applied to Romanian, with added Gaussian noise at different levels. The symbols $\uparrow$ and $\downarrow$ indicate that upper or lower values are better, respectively.}
\label{tab:noise}
\end{table}

\section{Conclusion and Future Work}

In this paper, we introduced a novel multilingual cross-domain audio dataset to evaluate deepfake detectors ``in the wild''. We discussed the methodology used to generate diverse fake clips for three partitions (training, in-domain test, and cross-domain test), and we further presented dataset statistics across the seven target languages. We evaluated the performance of six state-of-the-art models from the literature in terms of multiple metrics, showing that our cross-domain evaluation scenario causes a significant decline for all the reported metrics. While we were able to replicate the near perfect in-domain performance previously reported in literature, our cross-domain results highlighted the difficulty of performing audio deepfake detection across datasets, languages, speakers, and deepfake generative methods. 

In future work, we will focus on the development of robust domain adaptation techniques to improve the results in the cross-domain setup, which simulates a challenging real-world scenario.

\section{Acknowledgments}
This work was supported by a grant of the Ministry of Research, Innovation and Digitization, CCCDI - UEFISCDI, project number PN-IV-P6-6.3-SOL-2024-2-0227, within PNCDI IV. This research is also supported by the project ``Romanian Hub for Artificial Intelligence - HRIA'', Smart Growth, Digitization and Financial Instruments Program, 2021-2027, MySMIS no.~351416.

\section{Limitations}

To construct our audio deepfake detection benchmark, we relied on recent and publicly available text-to-speech and voice conversion methods. Unfortunately, most existing TTS and VC methods do not offer support for all the targeted languages. Hence, we were forced to use distinct generative methods across the chosen languages. Nevertheless, we kept in-domain vs.~cross-domain separation across all languages.

Our benchmark provides a challenging cross-domain evaluation setup, where noticeable performance drops can be observed. Yet, we did not try to adapt deepfake detectors to the cross-domain evaluation setting. Adapting models for the cross-domain setting is a challenge that requires careful consideration. We believe this exploration is beyond the goal of constructing a challenging benchmark for audio deepfake detection, so we leave it for future research. 

\section{Potential Risks}

The development of audio deepfake generation models can have significant implications for our society, as it facilitates the spread of misinformation and phishing attacks. As synthetic audio becomes increasingly realistic and accessible, the risk of misuse continues to grow. To fight against this, more competent detection models are required. Challenging dataset construction represents one way to advance research on robust detection systems, as such models heavily depend on the utilized training data. Our benchmark fosters the development of audio deepfake detectors, as it addresses some of the limitations of previous datasets: a variety of generation methods and languages, as well as a meticulously designed cross-domain split. Nevertheless, we acknowledge that the development of detection methods can inadvertently push research towards more sophisticated generative models. 

\section{License}

We share XMAD-Bench under the International Attribution Non-Commercial Share-Alike 4.0 (CC BY-NC-SA 4.0)\footnote{\url{https://creativecommons.org/licenses/by-nc-sa/4.0/deed.en}} license, aiming for open and responsible research on deepfake detection.

\bibliography{mybib}

\appendix

\end{document}